\documentclass[conference]{IEEEtran}
\IEEEoverridecommandlockouts
\usepackage{cite}
\usepackage{amsmath,amssymb,amsfonts}
\usepackage{algorithmic}
\usepackage{graphicx}
\usepackage{textcomp}
\usepackage{authblk}
\usepackage{xcolor}
\def\BibTeX{{\rm B\kern-.05em{\sc i\kern-.025em b}\kern-.08em
    T\kern-.1667em\lower.7ex\hbox{E}\kern-.125emX}}

\usepackage{hyperref}
    
\begin{document}

\title{Affective Conversational Agents: Understanding Expectations and Personal Influences

\thanks{\hspace{.1cm}\textsuperscript{+}Corresponding author: javierh@microsoft.com.}}

\author[]{Javier~Hernandez\textsuperscript{+}}
\author[]{Jina~Suh}
\author[]{Judith Amores}
\author[]{\\Kael Rowan}
\author[]{Gonzalo Ramos}
\author[]{Mary~Czerwinski}

\affil[]{Microsoft Research, Microsoft, Redmond, USA}

\maketitle

\begin{abstract}
The rise of AI conversational agents has broadened opportunities to enhance human capabilities across various domains. As these agents become more prevalent, it is crucial to investigate the impact of different affective abilities on their performance and user experience. In this study, we surveyed 745~respondents to understand the expectations and preferences regarding affective skills in various applications. Specifically, we assessed preferences concerning AI agents that can perceive, respond to, and simulate emotions across 32 distinct scenarios. Our results indicate a preference for scenarios that involve human interaction, emotional support, and creative tasks, with influences from factors such as emotional reappraisal and personality traits. Overall, the desired affective skills in AI agents depend largely on the application's context and nature, emphasizing the need for adaptability and context-awareness in the design of affective AI conversational agents.
\end{abstract}


\section{Introduction}
The development of conversational AI agents has been marked by significant milestones that have transformed the landscape of human-computer interaction. From early pioneering systems like Eliza in the 1960s~\cite{weizenbaum1966}, which simulated conversation by using pattern matching and substitution techniques, to the complex, context-aware agents that dominate today's technology ecosystem~\cite{hutson2018artificial}. This progression has been shaped by a rich history of research~\cite{turing1950,searle1980}, as well as popular culture, where science fiction movies and literature have envisioned a future where human-like machine interactions are seamlessly integrated into our lives~\cite{asimov1950}. As AI conversational agents continue to advance and permeate various domains, understanding their implications and potential becomes increasingly important for researchers, developers, and users alike.  
 
Text-based large language models have recently emerged as a major driving force in the field of AI conversational agents, demonstrating unique capabilities that set them apart from their predecessors. These models excel at encoding semantics, allowing them to communicate naturally with people and generate contextually relevant responses~\cite{vaswani2017}. Furthermore, they exhibit emergent behaviors~\cite{
bubeck2023sparks}, which stem from their ability to learn from massive amounts of textual data~\cite{radford2019}. Examples of such models include BERT~\cite{devlin2018}, which has been employed for various natural language processing applications such as sentiment analysis and question-answering systems~\cite{sun2019}, and more recently, OpenAI's GPT-4, which has shown proficiency in tasks ranging from text summarization to code generation~\cite{openai2023gpt4}. These models can operate as chatbots that can be customized to take different personas and skills, such as having the ability to recognize emotions associated with different perspectives, adjust their responses accordingly, and generate their own simulated states, enabling the development of more human-like and emotionally intelligent AI systems when appropriate. 

Existing literature highlights simulated affective empathy or emotional empathy as one important factor in these human-agent interactions~\cite{brave2005computers, leite2014empathic, paiva2017empathy}.
Affective empathy, as one of the core areas of the field of Affective Computing\cite{picard1997}, is one component of general empathy, which involves emotions as well as the capacity to share and understand another's state of mind, with continuous exchanges between emotion and intention in dialogue~\cite{harrelson2020intention, welivita2021large}.
Incorporating capabilities that embody affective empathy into systems has been shown to impact user satisfaction, trust, and acceptance of AI systems, as well as facilitate more effective communication and collaboration between humans and AI agents~\cite{marsella2010computational, rodriguez2014development}.
As conversational AI agents are further adopted across various application domains, it becomes crucial to understand the potential role and expectations of affective empathy in designing and building AI conversational agents with simulated affective empathy.

Motivated by these considerations, this work seeks to answer the following research questions:
\begin{description}
\item[\textbf{RQ1}]{What are the specific preferences of individuals regarding the manifestation of affective AI conversational agents across a diverse range of applications, and how do these preferences differ based on the context and nature of the application? }

\item[\textbf{RQ2}]{What human factors such as personality traits and emotional regulation skills contribute to the variation in personal preferences for affective AI agents, and how do these factors influence the acceptance of affective qualities?}  
\end{description}

To address these questions, we conducted a survey with 745~respondents, exploring their preferences for AI agents that can perceive, respond to, and simulate emotions~\cite{picard1997}. Our findings indicate that most of the participants would like agents with affective skills but the desired level of affective intelligence, as well as the particular abilities~(perceived, respond to, and simulate) depends largely on the nature of the application and personal human factors.

The paper is organized as follows. First, we review prior work on AI conversational agents, examining their applications, characteristics studied, and potential ethical concerns. Second, we delve into the methodology, detail the survey design and provide descriptive statistics about the data. Third, we present our analysis of the results, including preferences and contributing factors. Fourth, we discuss the limitations and implications of our work, along with potential avenues for future research. Finally, we provide some concluding remarks and relevant ethical considerations.

\section{Related Work}
AI conversational agents have been extensively explored in various application areas, including search, education, health care, entertainment, customer service, and social support, to provide information, assistance, feedback, or companionship~\cite{picard1997, bickmore2005establishing, Powell2022ConversationalAI, Martinengo2022ConversationalAI, Kusal2022AIBasedCA, Osta2021OnlineHC, Simpson2020DaisyAF, Ghosh2020PracticalET}. A key challenge in developing these agents is whether or not to incorporate aspects of Affective Computing which enables agents to recognize, interpret, and respond to human emotions. This capability has been shown to enhance user experience, engagement, and satisfaction by allowing AI agents to establish and maintain social and emotional bonds with users~\cite{sharma2021towards, leite2013social, leite2011modelling, Chawla2021TowardsEA, Cuylenburg2021EmotionGA}. Designing empathetic agents, however, requires not only selecting and integrating appropriate modalities to convey empathy but also considering the conversational style that best aligns with user preferences~\cite{LukeJesse21, tannen1987conversational, thomas2018style,Chin2019ShouldAA}.   

AI agents often employ natural language and other modalities, such as gestures and facial expressions, to interact with humans. These agents can take various forms, ranging from humanoid to non-humanoid and even non-physical entities~\cite{bickmore2005establishing}. Numerous comprehensive studies have investigated different aspects of agent voice, indicating that humans apply similar expectations to AI agents as they do to other humans, but these expectations may depend on factors such as the agent's quality, age, and embodiment~\cite{seaborn2021voice, james2018artificial}. Research on empathic AI agents, including those modeling empathy and prosocial behavior, has led to the development of systems that detect and regulate learners' affective states, adapt to personality traits, elicit and express emotions, and facilitate social and emotional learning in various contexts~\cite{santos2016emotions, paiva2017empathy, yalccin2020modeling, rashkin2018towards, zhou2020design, paiva2017empathy}. However, it is crucial to note that people with different conversational styles and personality traits may evaluate expressive agents differently, emphasizing the need for further research on the role of conversational style and user preferences in empathetic agents' effectiveness~\cite{thomas2020expressions, leary2011personality, richendoller1994exploring, costa2014associations}.

Despite the progress made in Affective Computing and AI conversational agents, several challenges and ethical implications still need to be addressed. These include accurately detecting and expressing emotions by the agent, the risk of over- or under-empathizing emotions, and potential emotional manipulation, which can have consequences for privacy, accountability, and human dignity~\cite{LukeJesse21, bostrom2014superintelligence}. Furthermore, most research to date has focused on specific scenarios or domains, without systematically exploring user preferences and expectations regarding affective empathy across a diverse range of applications, user backgrounds, and contexts.

\begin{table*}[t!]
\small
	\setlength{\tabcolsep}{4pt}

	\caption[]{Affect in Conversational Agents}
  \begin{tabular}{|p{.13\columnwidth}|p{.55\columnwidth}|p{.6\columnwidth}|p{.65\columnwidth}|}
    \hline
\textbf{\textit{Ability}} & \textbf{\textit{Definition}} & \textbf{\textit{Positive Example}} & \textbf{\textit{Negative Example}}  \\ \hline

Perceive & AI conversational agents could recognize the user's emotions via input cues such as tone, word choice, and context. & The agent detects excitement from the user's message, "I'm so thrilled with the results of this project!" & The agent detects disappointment from the user's message, "I'm really bummed that things didn't work out as planned." \\ \hline

Respond & In the future, agents could adjust their behavior and responses based on the user's emotions, expressing empathy or offering support. & The AI enthusiastically replies, "Congratulations on the amazing results! Your hard work has paid off. What's the next step for this project?" & The AI compassionately replies, "It is too bad that things didn't go as planned. It's understandable to feel disappointed. How can I help you move forward?" \\ \hline

Simulate & Future agents could generate and express their own simulated states to enhance interactions with users. & The AI simulates experiencing joy, saying: "Your success brings me joy too! I'm thrilled to be a part of your journey and support your continued growth." & The AI simulates experiencing sadness, saying, "I feel saddened to learn about the challenges you're facing. Please know that I'm here to help you navigate through these tough times and find a way to overcome them."\\ \hline

\end{tabular}
\label{table:ai_abilities}
\end{table*}

\section{Methodology}
This section outlines our approach to investigating affective empathy in AI conversational agents which includes the details surrounding the survey design and data collection.

\subsection{Survey Design}
To explore people's perspectives on affective empathy in AI conversational agents, we designed a survey with the following sections:

\textbf{Demographics and background information.} We collected basic demographic information about the participants, including their gender, age, job role, education level, and whether English was their first language.

\textbf{Personality.} To examine the role of personality on personal preferences, we included the 10-item Big Five Personality Inventory~\cite{rammstedt2007measuring} which decomposes personality into five main components: openness, conscientiousness, extroversion, agreeableness, and neuroticism.

\textbf{Emotional Experience.} To assess various aspects of participants' emotional experience, we included the 20-item Toronto Alexithymia Scale~\cite{bagby1994twenty} which provides an overall score for the inability to recognize or describe one's own emotions~(a.k.a.,~alexythmia) as well as separate scores for the difficulty to describe and identify feelings, as well as the ability for externally oriented thinking. We also included the 10-item Emotion Regulation Questionnaire~(ERQ)~\cite{gross2003individual} which provides scores for cognitive reappraisal and expressive suppression. These measures were selected based on their relevance to emotional intelligence, communication, and empathy, which are key components in the context of AI conversational agents.

\textbf{Prior experience.} To help contextualize the findings of this work, we evaluated participants' familiarity with the technology by asking them to rate the duration and frequency of their interactions with conversational AI agents, as well as describe some of their primary use cases.

\textbf{Preferences}. To capture preferences on affective empathy for AI conversational agents, participants were asked to indicate whether they would like to interact with an AI agent with simulated affective empathy to achieve different goals. Based on the seminal book of Affective Computing~\cite{picard1997}, we divided digital affective skills into three main components: the ability to perceive emotions from the user, the ability to respond to users' emotions, and the ability to simulate its own emotions. For each of these abilities, we provided a definition and some examples~(see Table~\ref{table:ai_abilities}). These dimensions were chosen to cover some of the core aspects of affective empathy~\cite{picard1997}, ensuring that our measures provided a nuanced understanding of participants' preferences in AI conversational agents. For each of these abilities, participants were then asked to rate whether they would like to see them when interacting with an AI agent across 32 application areas. 

\begin{figure*}[t!]
  \centering
  \includegraphics[width=.85\linewidth]{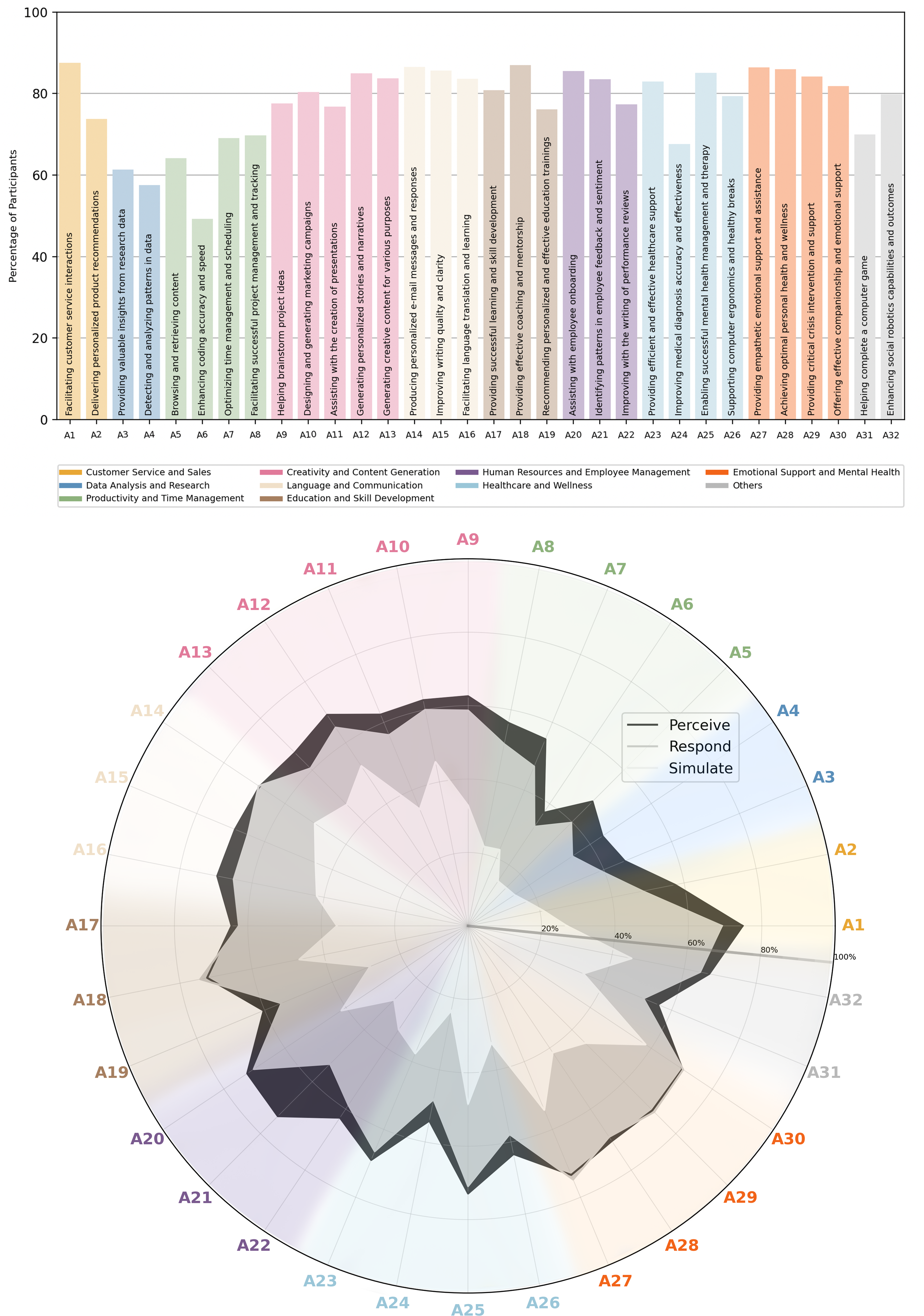}
  \caption{Overall preference for Affective AI conversational agents across different application areas~(top) and affective skills~(bottom).}
  \label{fig:bar_gradial}
\end{figure*}

We selected the list of application areas based on popular applications of text-based chatbots~(e.g.,~browsing and retrieving content, generating personalized stories and narratives), as well as prevalent application areas of emotion recognition~\cite{hernandez2021guidelines}~(e.g.,~providing empathetic emotional support and assistance, facilitating customer service interactions). These applications can be grouped into 10 main thematic categories~(e.g.,~customer service and sales, creativity, and content generation) as shown on Figure~\ref{fig:bar_gradial}~(top). 

Finally, to better understand potential individual variations in terms of affective preferences across all applications, we defined a Preference for Affective Conversations score~(PAC) for each participant. We computed this score as follows:
\begin{equation}
PAC_{p} = \frac{\sum_{i=1}^{n} I_{ip}}{n} \times 100\%
\end{equation}

where $n$ is the total number of applications, and $I_{ip}$ is an indicator variable that takes the value 1 if participant $p$ indicated at least one empathy preference for the $i^{th}$ application, and 0 otherwise.

\textbf{Open-ended scenarios.} Since the curated list of application areas might not have captured all possibilities, we asked participants to identify the ideal and worst application scenarios for an AI agent with affective empathy. For the ideal scenario, participants also provided information about the signals they would be willing to share with the AI conversational agent, as well as their preferences regarding the platform and potential channel of communication. To analyze open-ended questions, we conducted a thematic analysis regarding potential areas of application and concerns about affective AI agents. Each survey response was manually coded according to the themes we discovered to facilitate quantitative analysis.

\subsection{Data Collection and Overview}
A link to the survey was sent via email to a random set of information workers of a large technology company in the USA to request their preferences in terms of affective AI agents. Participation was voluntary without any form of compensation, and completing the survey took approximately 15~minutes. The study was pre-approved by the Institutional Review Board.

A total of 745~information workers took part in the user study, representing diverse job roles, education levels, age groups, and gender identities. The majority of participants were in the software development/engineering field~(55.84\%), followed by product management~(15.57\%), and data science/analytics~(3.62\%). Other job roles accounted for 21.36\% of the participants. The participants' education levels were predominantly bachelor's degrees~(46.44\%) and master's degrees~(35.30\%). A smaller percentage held other education levels, with a few preferring not to disclose their education~(1.34\%). In terms of age distribution, the largest group of participants fell within the 26-35 age range~(32.89\%), followed by those aged 36-45~(28.72\%) and 46-55~(20.81\%). A smaller percentage belonged to other age groups~(1.74\%). Gender-wise, the majority of participants identified as men~(70.20\%), while 25.64\% identified as women, and the remaining 1.47\% identified as non-binary/gender diverse, self-described, or preferred not to disclose their gender.

\section{Results}
This section presents our findings on various aspects of affective empathy, exploring preferred application areas, differences across preferred affective skills, and factors that influence potential acceptance.

\subsection{What prior experience does our sample population have with AI conversational agents?}

Prior experience with agents can play a role in how people perceive and understand interactions with agents. In our survey, 56.24\% of the participants reported having less than 6 months of experience with AI conversational agents, followed by 13.56\% with 6 months to 1 year, 8.86\% with 1-2 years, and 9.13\% had more than 5 years of experience. Regarding the frequency of interaction, 6.58\% of participants had never interacted with AI conversational agents, 15.57\% had interacted rarely~(once or twice in their life), 37.45\% interacted occasionally~(a few times a year), 22.68\% interacted frequently~(several times per month), and 17.72\% interacted very frequently~(multiple times per week or more). This highlights that most of the participants had at least occasional interactions with AI conversational agents.

Analysis of the free-form question regarding frequently explored applications revealed a variety of settings where respondents have used conversational AI agents, including customer service, personal assistance, healthcare, education, finance, e-commerce, entertainment, and writing, among others. Overall, these responses indicate that conversational AI agents are already widely used in numerous domains, suggesting a growing reliance on AI technology for various tasks.

\subsection{What are the most preferred application areas for incorporating affective empathy?}

Figure~\ref{fig:bar_gradial}~(top) shows the percentage of participants who indicated that they would like to see some kind of affective empathy, which includes any of the three abilities~(i.e.,~perceiving, responding to, or simulating emotions). As can be seen, there is large variability across applications with some consistency within categories. For instance, only around 50\% of the participants expressed a preference for \textit{affective empathy when enhancing coding accuracy and speed}~(A6), but 84\% favored its incorporation when \textit{producing personalized email messages and responses}~(A14). Interestingly, around 80\% of the respondents expected to see some form of emotional skills in AI agents for a large number of applications~(50\%), 

The most preferred categories for affective empathy were found to be \textit{language and communication}, \textit{emotional support and mental health}~(84.5\%~$\pm$~1.5\%), and \textit{creativity and content generation}~(80.8\%~$\pm$~3.2\%). These categories align with the inherently human aspects of interaction, empathy, and creativity, where the presence of affect in AI agents could be perceived as more beneficial. Conversely, the least preferred categories for affective empathy were \textit{data analysis and research}~(59.5\%~$\pm$~1.5\%), as well as \textit{productivity and time management}~(63\%~$\pm$~8.6\%). In these categories, participants may prioritize the efficiency and accuracy of AI agents over agents' ability to demonstrate affective skills, viewing affective empathy as less essential to the tasks at hand.

As part of the survey, we incorporated open-ended questions to explore potential areas where affective AI agents might be advantageous or disadvantageous. The thematic analysis of beneficial scenarios revealed preferences for mental health and emotional support~(27.7\%), healthcare and wellness~(21.9\%), customer service and experience~(20.7\%), coaching and parenting~(20.1\%), content creation and language education/translation~(10.7\%), personal assistants and productivity~(9.3\%), entertainment and gaming~(3.9\%), and personalized recommendations~(3.8\%). For the most prominent theme on mental well-being and support for individuals, respondents emphasized applications that assist individuals in various social contexts, such as therapy and counseling, as well as applications that offer companionship for those experiencing loneliness, isolation, or challenging life situations.

The thematic analysis of non-beneficial cases uncovered a mix of less-than-ideal scenarios and general concerns related to the concept of affective empathy. The distribution of themes included scenarios that require a high level of objectivity~(32.5\%), those necessitating emotional support and healthcare in general~(18.6\%), situations demanding emotional authenticity and human interaction~(17.2\%), general privacy and security concerns associated with emotional data~(7.5\%), marketing and customer service scenarios due to potential manipulation~(5.8\%), and general opposition against emotional capabilities~(5.5\%). For the most prominent theme of objectivity, respondents provided examples that were primarily technical, data-driven, or objective, such as coding, data analysis, and research. 

\begin{figure*}[t!]
  \centering
  \includegraphics[width=.9\linewidth]{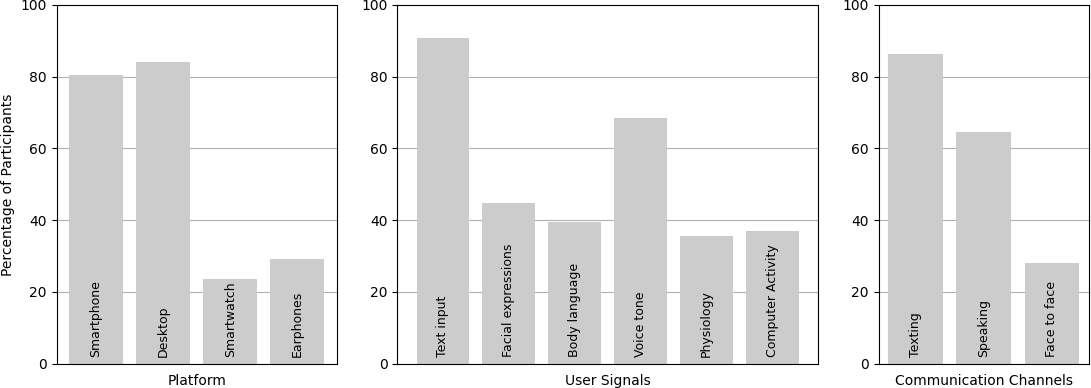}
  \caption{Preferences in terms of platforms~(left), shared user signals~(center), and communication channels~(right) for affective AI conversational agents.}
  \label{fig:pref_interactions}
\end{figure*}

\begin{figure}[t]
  \centering
  \includegraphics[width=.9\linewidth]{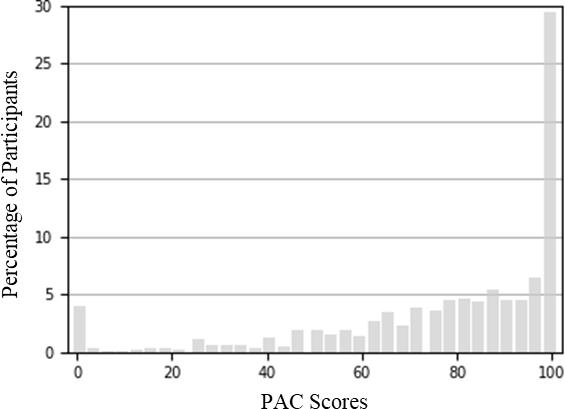}
  \caption{Distribution of Preference for Affective Conversations~(PAC) scores across participants. Higher PAC values indicate stronger preference towards seeing more affective empathy on AI conversational agents. }
  \label{fig:scores_hist}
\end{figure}

\subsection{How do preferences vary across affective skills and application areas?}

Figure~\ref{fig:bar_gradial}~(bottom) shows the percentage of users who preferred to interact with AI agents having the ability to perceive~(black), respond to~(grey), and simulate~(white) emotions across all the applications. The graph reveals that people generally expect AI agents to possess the ability to perceive emotions~(66.33\%~$\pm$~12.07\% on average), followed closely by the ability to respond to emotions~(61.87\%~$\pm$~12.86\%). Approximately half of the respondents expect AI agents to be capable of simulating emotions~(35.47\%~$\pm$~12.45\%), although there is noticeable variance across settings. For certain applications, such as \textit{producing personalized email messages and responses}~(A14), \textit{providing effective coaching and mentorship}~(A18), \textit{providing critical crisis intervention and support}~(A29), and \textit{offering effective companionship and emotional support}~(A30), the scores for perceiving and responding to emotions were nearly identical. 

Participants expressed a preference for AI agents that perceive and respond to emotions in applications involving human interaction or emotional support. Conversely, they expressed a preference for agents that can simulate emotions in creative or content generation tasks and emotional support. A lower preference for affective empathy, particularly for simulating emotions, was observed in data-focused or analytical tasks. Overall, the desired level of affect in AI agents appears to be highly dependent on the context and nature of the application. This underscores the importance of tailoring AI agent capabilities to meet the specific requirements and expectations of users across various domains.

When examining the open-ended questions, respondents shared diverse perspectives on the expected capabilities of affective AI agents. They acknowledged the potential benefits of AI agents that can accurately perceive and interpret users' emotional states, but also expressed concerns about privacy, data security, and the possibility of misinterpretation or emotional biases. Respondents also highlighted the advantages of AI agents capable of simulating emotional expressions to create more engaging, human-like, and relatable interactions. However, they raised concerns about authenticity and sincerity, as well as the potential for simulated empathy to be perceived as manipulative, disingenuous, or even harmful. They worried that AI agents could exploit users' emotional vulnerabilities. Interestingly, the contrasting views of benefits and concerns in areas such as mental health, emotional support, and customer service demonstrate the intricate and nuanced nature of public sentiment toward affective AI agents.

\subsection{What are the preferred methods of communication with affective AI agents?}

When evaluating preferences, respondents were asked to consider the application in which affective empathy made the most sense to them. Considering this application, Figure~\ref{fig:pref_interactions}~(left) shows the preferred platform for interacting with AI agents. In this case, participants showed a significantly higher preference for desktop platforms and smartphones than other devices such as smartwatches and earphones~(\(\chi^2(1)~=~938.57\), \(p < 0.001\)). Similarly, Figure~\ref{fig:pref_interactions}~(center) shows the signals that participants would be happy to share to enable affective empathy as well as the percentage of participants who selected them. As can be seen, the results indicate a significant preference for sharing text and voice data over other modalities such, as facial expressions, body language, and physiological signals~(\(\chi^2(1)~=~655.14\), \(p < 0.001\)). Finally, Figure~\ref{fig:pref_interactions}~(right) shows the preferred communication channels. As can be seen, people preferred significantly more text-based interactions~(messages) and audio~(speaking) interactions than face-to-face interactions~(\(\chi^2(1)~=~465.36\), \(p < 0.001\)).  

These findings show some important preferences in terms of platforms, user signals, and communication channels when designing affective AI agents. Our survey did not allow us to delve much deeper into why these preferences emerged, as we were unaware of them before this survey. We intend to follow up with future research on device/signal preferences. Still, by being aware of and potentially accommodating these preferences, developers can create more preferred, engaging, and affective agents in the future.

\subsection{How do emotional experience and personality traits impact acceptance of affective empathy?}
Figure~\ref{fig:scores_hist} shows the distribution of PAC scores across all participants, revealing large variance among individuals. For instance, around 30 participants did not select any affective skill across all applications, whereas around 220 people selected at least one empathy preference for all the applications.

To understand whether personal factors could explain PAC scores, we performed a linear mixed effects models in which we used the PAC scores as the dependent variable, and both personality and emotional experience information as the independent variables. All dependent variables were z-scored 
to ensure that a consistent scale was met across them. Table~\ref{tbl:lmm_results} shows the coefficients and statistical values, showing that multiple variables significantly contributed to the preference score. In particular, individuals with higher cognitive reappraisal skills contributed more significantly to a higher PAC score, suggesting that people who are able to reappraise their feelings effectively are more likely to prefer affective empathy. Similarly, higher ratings in difficulty identifying feelings decreased PAC scores, although less significantly.  In terms of personality, participants ranking high in extroversion and conscientiousness and, in a lesser amount, agreeableness, contributed positively to the score. This finding indicates that people who score high on these factors tend to prefer affective empathy on AI agents.

Finally, we examined the correlation between PAC scores and the likelihood of expressing concerns on the open-ended questions. As expected, we found that the opposition to the agent having affective capabilities was significantly and negatively correlated with users' preferences for affective capabilities overall~\mbox{($r$=-0.36)}, similarly with perceiving~\mbox{($r$=-0.19)}, responding to~\mbox{($r$=-0.28)}, and simulating~\mbox{($r$=-0.24)} capabilities. Similarly, concerns about emotional authenticity and human interaction were significantly and negatively correlated with the preferences for simulating capabilities~($r$=-0.19). 

\begin{table}[t!]
\centering
\caption[]{Factors contributing to the Preference for Affective Conversations \\
(‘$^{.}$’:$p<0.1$, ‘$^{*}$’:$p<0.05$, ‘$^{**}$’:$p<0.01$, ‘$^{***}$’:$p<0.001$).
}
    \begin{tabular}{| l | c | c | l |}
    \hline
     & \textbf{\textit{Coefficient}} & \textbf{\textit{$t$-value}} & \textbf{\textit{$p$-value}} \\
         \hline        
        \textit{Intercept} & 0.000 & 0.000 & 1.000 \\ \hline
        \textit{\textbf{Cognitive Reappraisal}} & \textbf{0.125} & \textbf{3.273} & \textbf{0.001~$^{***}$}  \\ \hline
        \textit{Expressive Suppression} & -0.064 & -1.491 & 0.136  \\ \hline
        \textit{Difficulty Describing Feelings} & -0.020 & -0.260 & 0.795  \\ \hline
        \textit{\textbf{Difficulty Identifying Feelings}} & \textbf{-0.109} & \textbf{-1.850} & \textbf{0.065~$^.$}  \\ \hline
        \textit{Toronto Alexithymia Scale} & 0.132 & 1.385 & 0.167  \\ \hline
        \textit{\textbf{Extroversion}} & \textbf{0.109} & \textbf{2.919} & \textbf{0.004~$^{***}$}  \\ \hline
        \textit{\textbf{Agreeableness}} & \textbf{0.067} & \textbf{1.799 }& \textbf{0.072~$^{.}$}  \\ \hline
        \textit{\textbf{Conscientiousness}} & \textbf{0.095} & \textbf{2.542} & \textbf{0.011~$^{*}$}  \\ \hline
        \textit{Neuroticism} & -0.022 & -0.579 & 0.563  \\ \hline
        \textit{Openness} & 0.054 & 1.486 & 0.138  \\ 
    \hline
    \end{tabular}

\label{tbl:lmm_results}
\end{table}

\section{Discussion}
This study provides valuable insights into people's preferences regarding affective empathy in conversational AI agents across a variety of applications. We found that there is a wide range of empathy preferences depending on the application, with higher affective expectations for tasks that involve human interaction, emotional support, and creative or content generation tasks. In particular, \textit{assisting in the writing of personalized e-mails and messages}~(A14), \textit{facilitating customer service interactions}~(A1), and \textit{providing effective coaching and mentorship}~(A18) received the highest ratings. Moreover, we identified preferences in terms of platforms, sharing of sensing modalities, and potential channels of communication in scenarios where affective empathy may be helpful. Understanding these preferences can help researchers and practitioners identify areas of opportunity and prioritize potential research and development efforts, ultimately leading to more effective and engaging AI conversational agents.

Our work demonstrates that certain personal characteristics can influence users' preferences in terms of affective empathy. We found that factors such as cognitive reappraisal skills and personality traits, such as extroversion and conscientiousness, can significantly contribute to users' expectations of affective empathy in AI agents. Our research method preventing us from knowing why these particular personality traits and characteristics drive certain preferences, but much further work needs to be done to explore the relationship. Also, these findings open an opportunity to explore the automated recognition of user preferences based on their personal information, which could be relevant in adjusting agents' abilities and better matching user expectations. Ultimately, this personalized approach to AI agent design could help enhance user experiences and create more meaningful interactions.

As part of the study, we recruited participants from a large technology company. While this allowed us to recruit people with prior experience with AI conversational agents, it will be important to validate our findings against other demographics and cultures. Future studies should consider more diverse populations to capture the complexities of emotion, such as cultural differences and the wider range of user preferences. Additionally, further exploration of other factors that might influence affective preferences, such as culture or environment, could provide a more comprehensive understanding of users' expectations.

The main focus of this work was to increase our understanding of people's preferences on expected affect in AI agents. As such, we did not go into the actual implementation and experience of AI agents with affective skills. 
Further research is needed to assess the potential impact of AI agents with affective empathy on achieving the different goals, as well as to evaluate user satisfaction and agent effectiveness in real-world scenarios. Moreover, there is room for investigation methods to implement and assess different affective skills, such as using fine-tuning, acting-related prompts, or other novel approaches that balance affective skills with other agent functionalities.

\section{Conclusions}
This work increases our understanding about people's preferences regarding affective empathy in AI conversational agents across a variety of applications. By surveying 745~respondents, we characterized their expectations for affect, preferred channels of communication and user signals, as well as potential influences of personal information on these preferences. Our findings revealed a wide range of preferences and highlighted the importance of context and application type in shaping users' expectations. Our research sheds light on the evolving field of conversational AI agents and the role of Affective Computing to help foster more meaningful and empathetic interactions.

\section{Ethical Statement}
Integrating affective empathy into AI agents is expected to present complex challenges that could potentially lead to user harms, including forced conformity, emotional harm, manipulation, and invasion of privacy. These risks may vary depending on the degree of empathy and the application domain, making it essential to carefully consider them when developing mitigation strategies. Some potential mechanisms to address these risks involve promoting responsible communication and informed consent, as recommended by prior research~\cite{hernandez2021guidelines, stark2021ethics, ong2021ethical, cowie2015ethical}. When interpreting and applying the findings of this study, it is crucial to be aware of its limitations, such as the distribution of the studied population and differences when compared to other populations. By acknowledging these limitations and incorporating ethical considerations into the research process, we can work towards developing AI agents that not only purposefully exhibit affective empathy but also respect users' autonomy and well-being, fostering a more ethically robust and user-centric perspective in future research endeavors.

\bibliographystyle{IEEEtran}
\bibliography{main}

\end{document}